\documentstyle[12pt]{article}
\begin{document}
\def\kms{km s$^{-1}$} \def\vlsr{$V_{\rm lsr}$} \def\Msun{M_\odot}
\def\sub{\subsection}

\title{Magnetic-Reconnection and Current-Sheet Model for the Radio Arc and Threads in the Galactic Center}
\author{Yoshiaki S{\sc ofue}\\
{\it Institute of Astronomy, University of Tokyo, Mitaka, Tokyo 181-0015} \\
Hiromitsu K{\sc igure}, and Kazunari S{\sc hibata}  \\
{\it Kwasan and Hida Observatory, Yamashina, Kyoto 607-8471}}

\maketitle

\abstract{ 
We propose a new mechanism to explain the radio Arc and threads in the Galactic Center by current sheets produced by local magnetic shears due to interaction of a moving cloud and the vertical field based on three-dimensional magneto-hydrodynamical simulations. Magnetic reconnetion and acceleration of cosmic-ray electrons in the current sheet will result in high contrast of radio emissivity inside and outside the Arc and threads.
}

\section{Introduction}

The Galactic Center is full of spectacular radio features, among which the nonthermal radio Arc is the most prominent unusual structure, believed to be highly ordered magnetic field vertical to the galactic plane. Each magnetic tube is as thin as one parsec or less and as long as a few tens of parsecs (Yusef-Zadeh, Morris, M., Chance 1984; Yusef-Zadeh, Morris 1987; Morris, Yusef-Zadeh 1985). Besides the Arc, there are numerous vertical radio filaments, called radio threads, also believed to be of magnetic origin (Anantharamaiah, et al. 1991; Lang, et al. 1999; LaRosa, et al. 2000; LaRosa, et al. 2004). 

Magnetic fields, as strong as $0.01 - 1$ mG, have been inferred for the radio Arc and threads from their large Faraday rotation measures (Lang, et al.1999; Tsuboi, et al.1986; Sofue, et al. 1987; Yusef-Zadeh, et al. 1997). Their straight structures indicate that they are composed of highly ordered magnetic fields perpendicular to the galactic plane. Although large-scale vertical magnetic fields appear to be common in the central regions of spiral galaxies (Tsuboi, et al. 1986; Sofue, et al. 1987; Sofue,  Fujimoto 1987), it has been a mystery why the Arc and threads are shining so locally, exhibiting high contrast of brightness to the ambient regions. 

A couple of models have been proposed for their origin. In a spark model for the radio Arc (Sofue, Fujimoto 1987), ejection of magnetized gas from Sgr A, observed as the thermal filaments, hit the vertical field in galactic rotation, and reconnection of the fields in the Arc and thermal filaments results in acceleration of cosmic-ray electrons to radiate nonthermal radio emission. On the other hand, a galactic wind model  (Dahrlburg, et al. 2002) suggests that interaction of the wind from the nucleus with a gas cloud causes wake instability, which grows to produce filamentary structures perpendicular to the galactic plane, mimicking the radio threads. The former theory, however, cannot explain the number of threads. The latter theory may explain the morphology of the threads, but cannot be applied to the strong Arc feature and is also difficult to eplain the high contrast of radio brightness in the threads.

In this letter, we propose a new mechanism to explain the Arc and threads  by current sheets produced by local magnetic shears due to ineteraction of a moving cloud and the vertical field. Magnetic reconnetion and acceleration of cosmic-ray electrons in the current sheet will result in high contrast of radio emissivity inside and outside the Arc and threads.

\section{Three-Dimensional Non-axisymmetric MHD Simulation of Cloud-Magnetic Field Interaction}

Vertical magnetic fields are often observed in spiral galaxies, including the Milky Way, and will be a common structure in the central regions (Sofue, Fujijmoto 1987). If there exists an accreting gas disk, the vertical field lines are twisted, and form a large-scale magnetic jet, and various simulations have been made in axisymmetric scheme (Uchida, et al. 1985; Uchida, Shibata 1985; Shibata, Uchida, Y. 1986, 1987). 

On the other hand, the distribution of molecular gas in the Galactic Center is far from axisymmetric, but is significantly shifted to positive galactic longitudes, and is very clumpy (Bally, et al. 1987; Oka, et al. 1998). Longitude-velocity diagrams also show non-axisymmetric kinematics, which is partly induced by accretion of gas by a bar structure (Sawada et al. 2004) . Such non-axisymmetric motions of gas will cause significant local disturbances of the magnetic fields, which may only be treated by a three-dimensional MHD calculations.

In order to examine how a locally moving gas cloud disturbs the ordered magnetic field, we performed three-dimensional MHD simulations. The calculations have been carried out using the CIP-MOC-CT scheme (Evans,  Hawley,  1988;  Yabe, Aoki, 1991; Yabe, et al.  1991; Stone,  Norman 1992; Kudoh, et al. 1999; Kigure, et al.  2004). The CIP scheme can accurately follow the contact discontinuity, or the surface of the cloud in the present case. This scheme has been developed to three-dimensional cylindrical code, and is now applicable to local disturbances of field lines near the origin of the coordinates (Kigure, H., and Shibata, K. 2005 in preparation). The numbers of meshes are $ 171 \times 100 \times 480$ in the radius, azimuth, and height $(r, \phi, z)$ directions.

We assume that the ambient magnetic field is at rest with respect to the rest coordinate frame, and the filed lines are perpendicular to the galactic plane defined by $(r,\phi, z=0)$.  The gravitational force was added by putting a point mass $M$ at the Galactic Center coinciding with the origin of the coordinates, which mimics the central massive core (Takamiya, Sofue 2000)  including the black hole  (Genzel, et al.  2003;  Ghez, et al. 2005). We put a gaseous corona around the central mass at hydrostatic equilibrium, but neglect its rotation.

The initial field strength is taken to be uniform at $B=B_0$. A gas cloud with a radius $r_{\rm w}$ is put at an initial position of $(r, \phi,z)=(1, 0, 0)$. The cloud is given of a Keplerian initial velocity $(v_{r}, v_{\phi}, v_{z})=(0, 1, 0)$. The gas cloud is assumed to have an initial density distribution represented by\begin{equation}
\rho = \rho_0 [ 1 + \cos ( {\pi | {x} - {x}_0 |}/{r_{\rm w}})]/2
\end{equation}
if $|{x} - {x}_0| < r_{\rm w}$ with ${x_0} = (r,\phi,z) = (1,0,0)$,
and we took $r_{\rm w} = 0.3$.
The density distribution of the corona around the center is given by
\begin{equation}
\rho = \rho_{\rm c} \rm{exp} [ \alpha \{ {r_0}/{(r^2 + z^2)^{1/2}} - 1\} ],
\end{equation}
where $r_0$ is the unit length. The parameter $\alpha$ is defined as $\alpha=\gamma V_{K0}^2 /s_{c}^2$, where $s_{c}$ is the sound velocity in the corona, $V_{K0} = (GM/r_0)^{1/2}$ is the Keplerian velocity at radius $r=r_0$. Here, $\rho_{\rm c}$ is the coronal density at radius $r_0$, $\alpha=1.0$ and $\rho_{\rm c}/\rho_0=10^{-3}$, where $\rho_0$ is the initial cloud density at the cloud center $(r, \phi, z) = (1, 0, 0)$.

The plasma-$\beta$, the ratio of gas pressure to magnetic pressure defined by $\beta=2s^2/\gamma V_{\rm A}^2$, is  taken to be 0.4 in the cloud center, and is of the order of $\sim 10^{-2}$ in the ambient coronal region, where $\gamma=5/3$ is the specific heat ratio of the gas. Here, $s$ is the sound velocity at the clod center. We assume a free boundary condition. The gravitational and kinetic energy densities in the cloud are much greater than the magnetic field energy density. Hence, the field lines are nearly passive to the motion and distortion of the cloud. Initially we take $u_{\rm m}/u_{\rm g} \sim u_{\rm m}/u_{\rm k} =(V_{A0}/V_{\rm K0})^2=5 \times 10^{-3}$, where $V_{\rm A0}$ and $V_{\rm K0}$ are the Alfven velocity and the velocity of cloud. Here, $ u_{\rm g} = G M_0 \rho /r$, $u_{\rm k} = \rho v^2 /2$, and $u_{\rm m} = B^2/8\pi $. On the other hand, we have $(V_{\rm Ac}/V_{\rm k0})^2  \sim 5$ for the coronal gas, and therefore, the magnetic field is not strongly disturbed by the corona. The gas pressures between the corona and cloud are assumed to be balanced, so that $ (s_{\rm c}/s)^2 =\rho_0/\rho_{\rm c} \sim 10^3$, where $s_{\rm c}$ is the sound velocity in the corona.

Figure 1 shows the result of the three-dimensional MHD simulation. Vertical lines are magnetic lines of force. The $(x,z)$ plane is indicated by a grey-colored panel, with the grey brightness being proportional to the coronal gas density. A gas cloud is rotating around the center in Keplerian motion.  The top panels show the time evolution of field lines from $t=0$ to 2, and the bottom-left panel is at $t=3$. Here, time $t$ is normalized by $P/2\pi$ with  $P$ being the rotation period of the cloud around the center. As the cloud moves, the magnetic field lines are wound, and twisted locally due to the cloud motion as well as the gravitational distortion of the cloud shape. Note that the viewing point of figure 1 is rotating with the cloud motion, so that the $(x,z)$ plane is apparently rotating in the figure.

\centerline{ --- Fig. 1 ---}

According to the twisting distortion of the field lines, a current sheet is produced, along which magnetic reconnection is supposed to take place. The bottom-left panel of figure 1 shows an equal-current-density surface with $J=1.5$ at $t=3$. Here, $J$ is the current density calculated by $J={\rm rot}B$, and is normalized by $J_0=b_0/r_0 = [\rho_0 V_{k0}^2]^{1/2}/r_0$, where $b_0$ is the non-dimensionalized initial magnetic strength. We emphasize that the thus formed local enhancement of current density produces a feature mimicking the radio Arc and thread. The bottom right panel shows the radio Arc as reproduced from VLA observations at 21 cm (Yusef-Zadeh, Morris 1987).

\section{Magnetic-Reconnection and Current Sheet Model for Arc and Threads}

Based on the simulation, we propose a new model to explain the origin of radio Arc and threads in the Galactic Center, which we call the magnetic-reconnection and current-sheet (MRC) model.

Suppose a gas cloud with different velocity from that of the field lines hit the vertical magnetic lines of force, such as due to infalling cloud, ejection from the center, or shock waves. The cloud will locally twist the vertical magnetic field. Figure 2 illustrates how a moving cloud influences the magnetic field lines. The locally twisted field lines, then, produce a magnetic shear between the twisted and the ambient fields. This magnetic shear produces a current sheet, along which magnetic reconnection would occur. This mechanism is similar to the nano-flare model for solar-coronal heating (Parker 1988). In the present case, the twisted field looks more like a bunch of lines of force than a sheet, and, hence, the current sheet would be more like a ``current thread'',  as figures 1 and 2 show.

\centerline{--- Fig. 2 ---}

The flux density of the released energy per unit volume along the current sheet can be expressed by the Poynting flux,
$$ f \sim ( B_z B_\phi/4 \pi) V$$
where $B_z$, $B_\phi$, and $V$ are the magnetic strengths in the vertical and azimuthal directions, and the velocity of gas, respectively.

The vertical component $B_z$ is on the order of the initial value, $B_z \sim {B_z}_0$. On the other hand, the initial value of ${B_\phi}_0$ is almost zero, while it is created during the twist to attain a finite value $B_\phi$, which we denote as $B_\phi \sim \xi {B_z}_0$. Hence, the flux of released energy along the current thread is of the order of
$$ f \sim  \xi ({{B_z}_0}^2/4 \pi) V.$$
In the present numerical calculation, we have $\xi \sim 0.1$. This flux should be compared with the one in the ambient region,
$$ f_0  \sim ({B_z}_0 {B_\phi}_0/4 \pi) V_0.$$
Thus, the ratio of the released fluxes inside and outside the current thread is of the order of
$$ \eta \sim f/f_0 \sim \xi [{B_z}_0/ {B_\phi}_0][ V/V_0].$$
The ratio $\eta$ will attain an extremely large value, because the ambient region has almost negligible value of ${B_\phi}_0$, as there exists neither a magnetic twist nor a current sheet.

The energy released in the current sheet will be used to accelerate cosmic-ray electrons, which interact with the magnetic field and radiate synchrotron emission. We emphasize that the radio emissivity in the current sheet (thread) increases significantly even without large amplification of the field strength because of the large $\eta$ value. Let the fraction $\delta$ of the Poynting flux $f$ be transformed to the synchrotron emission. Then, we obtain the radio flux of
$$ f_{\rm r} \sim  \delta \xi ({{B_z}_0}^2/4 \pi) V.$$
The surface brightness of radio emission at frequency $\nu$ is, then, estimated to be of the order of
$$ \Sigma = f/\nu \sim \delta \xi B_0^2 V/4\pi \nu.$$
If we take $B_0 \sim 10^2 \mu$G, $V \sim 100$ kms s$^{-1}$,  and $\nu \sim 1$ GHz, we obtain $ \Sigma \sim \delta \xi 10^{-11}$ erg s$^{-1}$ cm$^{-2}$ Hz$^{-1}$, or $\Sigma \sim \delta \xi 10^{-14}$ W m$^{-2}$ Hz$^{-1}$. This leads to brightness temperature at 1 GHz of $T_b = \delta \xi \lambda^2 \Sigma/2k \sim 10^5 \delta \xi $ K at around $\nu \sim 1$ GHz, where $\lambda = c/\nu$ is the wavelength and $k$ is the Boltzmann constant.

The radio brightness for the radio Arc and threads has been observed to be of the order of  $T_b \sim 0.1$ K at $\sim 1$ GHz (Yusef-Zadeh, Morris 1987;  LaRosa et al. 2004). Here, $\xi$ is approximately $\sim 0.1$ from the simulation. The transformation efficiency $\delta$ of Poynting flux to radio emission is a subject to sophisticated treatments of magnetic reconnection and acceleration of cosmic rays. Instead, we, here, estimate a possible value of the paprameter $\delta$. In order for the formulated radio brightness to agree with the observed brightness, we require only a very small efficiency on the order of $\delta \sim 10^{-5}$.

We have discussed the MRC model for one particular cloud. However, due to barred galactic shocks as well as angular momentum loss by large-scale magnetic field twist, a number of clouds are infalling to the Galactic Center. Recurrent interaction of gas clouds and clumps will, therefore, result in many "threads" in the Galactic Center region, as indeed observed (LaRosa et al. 2000, 2004). Among the many threads, the strongest interaction between a cloud and magnetic field is presently observed at the radio Arc. In fact, the field lines in the Arc are interacting with a dense gas cloud (Tsuboi, et al. 1997). Interaction with the thermal filament extending from Sgr A with the Arc (Sofue, Fujimoto 1987) may also infer local twist. Such disturbances will be caused not only by accreting gas clouds, but also by hydrodynamical waves and shocks triggered by various activities in the Galactic Center region, such as the star formation, gas ejection from the center, etc.. Analyses of the shapes, distribution, and frequency of radio threads may give information about hydrodynamical conditions in the Galactic Center region.

Numerical computations were carried out on VPP5000 at the Astronomical Data Analysis Center of the National Astronomical Observatory, Japan (project ID: rhk05b and whk08b), which is an interuniversity research institute operated by the Ministry of Education, Culture, Sports, Science, and Technology.

\vskip 10mm

\noindent{\bf References}
\def\r{\noindent \hangindent=10pc}

\r Anantharamaiah, K. R., Pedlar, A., Ekers, R. 1991 MNRAS 249, 262

\r Bally, J., Stark, A., Wilson, R. W., Henkel, C.  1987 ApJS 65, 13

\r Dahrlburg, R. B., Einaudi, G., LaRosa, T. N., Shore, S. N. 2002 AJ 568, 220

\r Evans, C. R., Hawley, J.P. 1988, ApJ 332, 659.

\r Genzel, R., Schodel, R., Ott, T., Eisenhauer, F., Hofmann, R., et al.  2003 ApJ 594, 812

\r Ghez, A. M., Salim, S., Hornstein, S. D., Tanner, A., Lu, J. R., Morris, M., Becklin, E. E., Duchene, G. 2005, ApJ 620, 744

\r Kigure, H., Uchida, Y. Nakamura, M., Hirose, S., Cameron, R., 2004 ApJ 608, 119

\r Kudoh, T., Matsumoto, R., Shibata, K. 1999, Comput. Fluid Dyn. J. 8, 56.

\r LaRosa, T. N., Kassim, N. E., Lazio, T. 2000 AJ 119, 207

\r LaRosa, T. N., Nord, Michael E., Joseph, T.,  Lazio, W.,  Kassim, Namir E.  2004 ApJ 607, 302

\r Lang, C. C.,  Morris, M.,  Echevarria, L. 1999 ApJ 526, 727

\r Morris, M., Yusef-Zadeh, F. 1985 AJ 90 2511

\r Oka, T., Hasegawa, T., Sato, F., Tsuboi, M., Miyazaki, A. 1998 ApJS 118, 455

\r Parker, E. 1988, ApJ 330, 474.

\r Sawada, T., Hasegawa, T., Handa, T., Cohen, R. J., 2004 MNRAS 349, 1167.

\r Shibata, K., Uchida, Y. 1986, PASJ, 38, 631

\r Shibata, K., Uchida, Y. 1987, PASJ 39, 559 %GCL

\r Sofue, Y., Fujimoto, M. 1987 ApJ 319, L73

\r Sofue, Y., Fujimoto, M. 1987 PASJ 39, 848

\r Sofue, Y., W.Reich, M.Inoue, J.H.Seiradakis 1987 PASJ 39, 95

\r Stone, J. M., Norman, M. L. 1992, ApJS, 80, 791.

\r Takamiya, T., Sofue, Y. 2000 ApJ 534, 670

\r Tsuboi, M., Inoue, M.,  Handa, T., Tabara, H.,  Kato, T.,  Sofue, Y., Kaifu, N. 1986 AJ 92, 818

\r Tsuboi, M., Ukita, N., Handa, T., 1997 ApJ, 481, 263.

\r Uchida, Y., Shibata, K. 1985, PASJ, 37, 515

\r Uchida, Y., Sofue, Y., Shibata, K. 1985 Nature 317, 699

\r Yabe, T., Aoki, T. 1991, Comp. Phys. Comm. 66, 219.

\r Yabe, T., Ishikawa, T., Wang, P. Y., Aoki, T., Kadota, Y., Ikeda, F. 1991,  Comp. Phys. Comm. 66, 233.

\r Yusef-Zadeh, F., Morris, M., and Chance, D. 1984, Nature 310, 557.

\r Yusef-Zadeh, F., Wardle, M., Parastara, P. 1997 ApJ 475, L119

\r Yusef-Zadeh, F., and Morris, M. 1987, ApJ, 322, 721

\newpage

\parindent=0pt
\parskip 6mm

Figure Captions

Fig. 1. Three-dimensional, non-axisymmetric MHD simulation of interaction of a gas cloud with vertical magnetic field in the Galactic Center region. The field is embedded in a hydrostatic gaseous halo around a central mass. Upper panels show the evolution of magnetic lines of force at $t=0, 1$ and 3. The moving cloud is disturbed by the gravitation, and locally twists the magnetic field. For the units of time and linear scale, see the text.The lower-left panel shows an equal-current-density surface of the current sheet (current thread) at $J=1.5$ for $t=3$. The lower-right panel is the radio Arc (vertical structures) near the Galactic Center (the brightest spot) (Yusef-Zadeh,  Morris 1987).

Fig. 2. Schematic illustration of a scenario for the origin of Galactic Center radio threads, due to local twist of vertical magnetic field lines. Reconnection of the field lines along current sheets (threads) accelerates cosmic-ray electrons to emit synchrotron radiation. Inserted in the bottom-right corner is the observation of 90-cm radio threads (LaRosa et al. 2000).

\end{document}